\documentclass[prl,twocolumn,epsf,psfig]{revtex4}
\usepackage{graphicx}
\DeclareGraphicsExtensions{.pdf,.png,.gif,.jpg}
\usepackage{float}
\usepackage{subfig}
\usepackage{epsfig}
\usepackage{wrapfig}

\begin{document}

\title{Metal-Insulator-Superconductor transition of spin-3/2 atoms on optical lattices}

\author{Theja N. De Silva}
\affiliation{Department of Chemistry and Physics,
Augusta University, Augusta, GA 30912, USA.\\
Kavli Institute for Theoretical Physics, University of California, Santa Barbara, CA 93106, USA. \\
Institute for Theoretical Atomic, Molecular, and Optical Physics,Harvard-Smithsonian Center for Astrophysics, Harvard University, Cambridge, MA 02138, USA.}

\begin{abstract}
We use a slave-rotor approach within a mean-field theory to study the competition of metallic, Mott-insulating, and superconducting phases of spin-3/2 fermions subjected to a periodic optical lattice potential. In addition to the metallic, the Mott-insulating, and the superconducting phases that are associated with the gauge symmetry breaking of the spinon field, we identify a novel emerging superconducting phase that breaks both roton and spinon field gauge symmetries. This novel superconducting phase emerges as a result of the competition between spin-0 singlet and spin-2 quintet interaction channels naturally available for spin-3/2 systems. The two superconducting phases can be distinguished from each other by quasiparticle weight. We further discuss the properties of these phases for both two-dimensional square and three-dimensional cubic lattices at zero and finite temperatures.
\end{abstract}

\maketitle

\section{I. Introduction}

Recent extraordinary progress achieved in trapping and manipulating ultra-cold atomic gases provides a wonderful opportunity for exploring quantum many-body physics. Ultra-cold atomic systems are now considered as one of the  most promising and efficient playgrounds for studying condensed matter and nuclear physics phenomena~\cite{reviewTH}. Recent developments in laser technology and experimental advancements allow one to have unprecedented control over various experimental parameters~\cite{reviewEX}. Effective spatial dimensionality, lattice structure, and lattice geometry can be tuned by adjusting the laser intensity, phase, and wavelength. The interaction between the atoms can be controlled dramatically by adjusting the two-body scattering length through magnetically tuned Feshbach resonance. Through the first generation of experiments with ultra-cold bosons and fermions in optical lattices, it has been well established that these systems can exhibit a variety of interesting phenomena~\cite{FDop1, FDop2, FDop3, FDop4, FDop5, FDop6}. The growing availability of multi-component degenerate fermionic atoms, such as $^6$Li~\cite{Lisys1, Lisys2, Lisys3}, $^{40}$K~\cite{Ksys}, $^{135}$Ba and $^{137}$Ba~\cite{Basys}, and $^{173}$Yb~\cite{Ybsys} provides a controllable platform to study higher spin, strongly correlated physics that features novel phenomena.

Among multi-component ultra-cold gases, high spin fermions such as spin- 3/2 $^{132}$Cs, $^9$Ba, $^{135}$Ba, and $^{201}$Hg attracted much attention due to the rich collective phenomena they can exhibit~\cite{HSF1, HSF2, HSF3, HSF4, HSF5, HSF6, HSF7, HSF8, HSF9, HSF10, HSF11, HSF12}. Spin-3/2 systems are expected to show emerging behavior due to the competing parameters, such as \emph{total } spin-0 singlet and spin-2 quintet scattering lengths. The \emph{total} spin-1 and spin-3 channels are prohibited due to the Pauli exclusion principle. In addition, the strong quantum fluctuations due to the enlarged SO(5) or Sp(4) symmetry is expected to play a bigger role in these systems~\cite{Wu1, chen}. In particular, when these spin-3/2 atoms are subjected to a periodic lattice potential, they can show novel collective behavior that is not obvious in spin-1/2 electronic systems. For example, on-site four-particle clustering instabilities can leads to quintet Cooper pairing favored by the spin-2 interaction channel~\cite{FPC1, FPC2, FPC3, FPC4, FPC5, FPC6}. When the total spin-0 interaction channel is strongly positive, the system can leads to a Mott-insulating state with a fixed number of atoms on each lattice site. It is the purpose of this paper to study the competition and phase transitions among metal, Mott-insulating, and singlet Cooper pairing states of neutral spin-3/2 fermions subjected to a two-dimensional square lattice and a three-dimensional cubic lattice. In order to do so, we use a slave rotor approach that allows us to handle the intermediate coupling regime where the charge fluctuations are strong~\cite{florens}. In the slave rotor representation, the particle operator is decomposed into a roton-bosonic field and a spinon-fermionic field. While the roton caries the charge degrees of freedom, the spinon caries the spin degrees of freedom. In this approach, the metal and Mott-insulating phases are characterized by breaking of global U(1) gauge symmetry associated with the charge degrees of freedom. In general the superconducting phase is characterized by breaking of global U(1) symmetry associated with the spinon degrees of freedom.

\noindent In addition to the obvious metal, Mott-insulator, and conventional (in the sense that one gauge symmetry is broken) superconducting phases arising from the competing interactions, we find a novel emerging superconducting phase where both global symmetries associated with charge and spin degrees of freedom are broken. This novel superconducting phase is differentiated from the conventional superconducting phase due to the non-zero quasi particle weight. Notice that we use the condensed matter terminology, but our metal and superconducting phases are neutral for atoms in optical lattices. Further, we investigate each of these emerging phases at both zero temperature and finite temperature by calculating various physical quantities.

The paper is organized as follows. In section II, we introduce a spin-3/2 model Hamiltonian for atoms on an optical lattice. The model is a generalized Hubbard model based on microscopic s-wave atom-atom interaction. In section III, we introduce the slave rotor approach and convert our model Hamiltonian in to a coupled rotor-spinon Hamiltonian. In section IV, a decoupling scheme is introduced to decouple the rotor part and the spinon part of the Hamiltonian. In section V and VI, we use a mean-field treatment to solve the rotor and spinon sectors of the Hamiltonian. In section VII and VIII, we discuss our zero temperature and finite temperature formalism and their results. In section IX, we summarize and discuss our results.

\section{II. Model Hamiltonian}

We start with the generic form of the spin-3/2 neutral particle Hamiltonian of the lattice model~\cite{Wu1},

\begin{eqnarray}
H = -t\sum_{\langle ij \rangle}(c^\dagger_{i\sigma}c_{j\sigma} + h.c)-\mu\sum_{i \sigma} c^\dagger_{i\sigma}c_{i\sigma} \nonumber \\
+U_0\sum_i P^\dagger_{00}(i)P_{00}(i) +U_2 \sum_{i, m=\pm 2,\pm 1,0} P^\dagger_{2m}(i) P_{2m}(i),
\end{eqnarray}

\noindent where $P^\dagger_{Fm} (i) = \sum_{\alpha \beta} \langle F, m|\alpha, \beta \rangle c^\dagger_{i \alpha} c^\dagger_{i \beta}$ are the singlet ($F =0$, $m = 0$) and quintet ($F = 2$, $m$) pairing operators and $c^\dagger_{i,\sigma}$ is the fermionic creation operator at site $i$, in one of the hyperfine spin state $\sigma = \pm 1/2, \pm 3/2$. Here,

\begin{eqnarray}
U_s = \int d\vec{r} d\vec{r^\prime} w^\ast(\vec{r}-\vec{R}_i)w^\ast(\vec{r^\prime}-\vec{R}_i) \\ \nonumber
\times g_s w(\vec{r^\prime}-\vec{R}_i) w(\vec{r}-\vec{R}_i),
\end{eqnarray}

\noindent is the interaction parameter for the total spin $S=0$ and $S = 2$ channels with the contact interaction in free space $g_s = 4 \pi \hbar^2 a_s/m$ and localized Wannier functions $w(\vec{r}-\vec{R}_i)$ at $\vec{R}_i$, where $a_s$ is the s-wave scattering length for the total spin-S channel. At half filling (ie, on average one atom per site), the particle-hole symmetry ensures the chemical potential $\mu = (U_0 + 5 U_2)/4$~\cite{Wu1}. Here we assume that the atoms can hop between nearest-neighbors with hopping amplitude $t$, where $\langle ij \rangle$ stands for the sum over only nearest neighbors.

For the purpose of studying the phase transition of metallic, insulating, and superconducting phases, it is convenient to re-write the Hamiltonian in terms of the spin-3/2 on-site singlet operator $P^\dagger_i \equiv P^\dagger_{00}(i) = 1/\sqrt{2}( c^\dagger_{i,3/2} c^\dagger_{i,-3/2}-c^\dagger_{i,1/2} c^\dagger_{i,-1/2})$ and the on-site density operator $n_i = \sum_{\sigma} c^\dagger_{i \sigma} c_{i\sigma}$~\cite{Roux},

\begin{eqnarray}
H = -t\sum_{\langle ij \rangle}(c^\dagger_{i\sigma}c_{j\sigma} + h.c)-\mu_0\sum_{i \sigma} c^\dagger_{i\sigma}c_{i\sigma} \nonumber \\
+U/2 \sum_i (\sum_{\sigma} c^\dagger_{i\sigma}c_{i \sigma}-4/2)^2 + V \sum_{i} P^\dagger_{i} P_{i}
\end{eqnarray}

\noindent Here $U = 2U_2$ and $V = U_0 - U_2$ with the shifted chemical potential $\mu_0$ is given by $(U_0-U_2)/4$ at half filling. This model has an exact SO(5) symmetry which reduces to a SU(4) symmetry at $U_0 = U_2$~\cite{Wu1, HSF2}. Notice that at this SU(4) symmetric limit, the chemical potential at half filling reaches zero. Each major terms in the model competes for metallic, Mott-insulating, and singlet pairing states. At the SU(4) limit and for interactions that are weak compared to the tunneling energy, $ U \ll t$, atoms can gain kinetic energy by hopping through the lattice. In the opposite limit where $U \gg t$, the repulsion is greater than the gain in kinetic energy, thus the atoms will localize at lattice sites, resulting in a Mott-insulator. For $V <0$, the model naturally favors the singlet pairing state while the first two terms compete for metallic and Mott-insulating states, respectively. In addition to the singlet pairing, as discussed in Ref.~\cite{HSF11}, multi-particle clustering of spin-3/2 atoms can lead to quintet pairing states with a total spin-2. The quintet pairing requires a negative quintet interaction parameter $U_2$. Here we consider a positive $U_2$ that supports Mott-insulating states, therefore we can safely neglect the possible quintet pairing in the model.

For deep optical lattices, one can approximate the Wannier functions by the Gaussian ground state in the local oscillator potential and find that the tunneling amplitude $t \propto E_r V_r^{3/4} e^{-2 \sqrt{V_r}}$ is exponentially sensitive to the laser intensity $V_0 = E_rV_r$ that was used to create the optical lattice, here $E_r$ is the recoil energy~\cite{bloch}. The interaction terms $U_s \propto a_s E_r V_r^{3/4}$ is linearly sensitive to scattering lengths and relatively, weakly sensitive to the laser intensity. As a result, the model is highly tunable in experimental setups in optical lattice environments.

\section{III. Slave-rotor approach}

Slave-particle approaches are proven to be simple and computationally inexpensive approaches to study strongly correlated effects in many-particle systems and these approaches are capable of accounting for particle correlations beyond standard mean-field theories. The first slave-particle approach was proposed to study the Mott insulator-metal transitions~\cite{GUT}. There are several advantages of using slave-particle approaches over other mean-field theories and variational methods. While most variational approaches are valid only at zero temperature, the slave-particle approaches are applicable at both zero and finite temperatures. Unlike other mean-field theories, quantum fluctuations can be taken into account by the Stratonovich-Hubbard transformation within the slave particle formalism~\cite{SHT}. Further, it has been shown that slave-particle approaches are equivalent to a statistically-consistent Gutzwiller approximation~\cite{GEV1, GEV2, GEV3}. Here we use the slave-rotor approach as it is convenient for many component systems~\cite{florens}. The method is simply to introduce an auxiliary boson to represent the local degrees of freedom in the correlated system. The metallic solution will be described as a correlated Fermi liquid. In the slave-rotor approach, the original local Fock space of the problem is mapped onto a larger local Fock space that contains as many fermions degrees of freedom as the original one and the same number of spin-3/2 local quantum variables, one for each fermion.  While a new pseudo-fermion variable describes the itinerant quasiparticle fraction of the atom, the auxiliary boson describes its localized fraction.

The slave-rotor approach was first introduced by Florens and Georges for the Hubbard model to study the metal-Mott insulator transition~\cite{florens}. Later this approach was applied to magnetic systems to study spin liquid phases~\cite{SLP1, SLP2, SLP3, SLP4, SLP5, SLP6}. In this approach, the particle operator is decoupled into a fermion and a bosonic rotor that carry the spin and the charge degrees of freedom, respectively. First, the particle operator $c_{i \sigma}$ that annihilates an atom with spin $\sigma$ at site $i$ is expressed as a product:

\begin{eqnarray}
c_{i \sigma} = e^{-i \theta_i} f_{i \sigma},
\end{eqnarray}

\noindent where the auxiliary fermion $f_{i \sigma}$ annihilates a spinon with spin $\sigma$ and the local phase degree of freedom $\theta_i$ conjugates to the total (neutral) charge through the "angular momentum" operator $L_i = -i\partial/\partial \theta_i$,

\begin{eqnarray}
[\theta_i, L_j] = i \delta_{ij}.
\end{eqnarray}

\noindent In this representation, while the rotor operator $e^{-i \theta_i}$ reduces the site occupation by one unit, the eigenvalues of the $L_i$ correspond to the possible number of atoms on the lattice site. Notice that the term "angular momentum" is used due to the conservation of the $O(2)$ variable $\theta_i \in [0, 2 \pi]$ but has nothing to do with the physical angular momentum of the atoms. Using the fact that rotons and spinons commute, one can show that the number operator of the physical particles coincide with that of the spinon;

\begin{eqnarray}
n_{i \sigma} = c^\dagger_{i \sigma} c_{i \sigma} = f^\dagger_{i \sigma} f_{i \sigma} = n^f_{i \sigma}.
\end{eqnarray}

\noindent As the eigenvalues of the angular momentum operator $l \in \textit{Z}$ can have any integer values, one must impose a constraint to truncate the enlarged Hilbert space to remove unphysical states,

\begin{eqnarray}
L_i = \sum_{\sigma} n^f_{i \sigma} - 1.
\end{eqnarray}

\noindent This constraint connects the charge and spin degrees of freedom and can be taken into account by introducing a Lagrange multiplier in the formalism. Notice that the angular momentum operator $L_i$ measure the particle number at each site relative to the half-filling. In terms of new variables, our Hamiltonian in Eq. (3) becomes,

\begin{eqnarray}
H = -t\sum_{\langle ij \rangle}f^\dagger_{i\sigma}f_{j\sigma} e^{i(\theta_i-\theta_j)}-(\mu_0 + h) \sum_{i \sigma} f^\dagger_{i\sigma}f_{i\sigma} \nonumber \\ +\frac{U}{2} \sum_i L_i^2 + + V \sum_{i} P^\dagger_{i} P_{i},
\end{eqnarray}

\noindent where pairing operators in the last term now has the form $P^\dagger_i = 1/\sqrt{2} (f^\dagger_{i,3/2} f^\dagger_{i,-3/2}-f^\dagger_{i,1/2} f^\dagger_{i,-1/2}) e^{2 i \theta_i}$. Notice that the constraint is treated on average so that the Lagrange multiplier $h$ is site independent. Even though one interaction term in the $S = 2$ channel simply becomes the kinetic energy for the rotons, the pairing interaction term is still quartic and the hopping term now becomes quartic in spinon and rotor operators as well. In the following section, we make further approximations to the quartic terms to get a manageable theory.

\section{IV. Decoupling spinon and rotors}

For spin-3/2  atoms on square or cubic lattices at half filling, we plan to decouple the Hamiltonian in Eq. (8) by using a mean field description. First, we decouple the hopping term so that the Hamiltonian $H$ becomes the sum of independent spinon and rotor parts: $H \rightarrow H_f + H_\theta$. This will lead to the $H_\theta$ part being an interacting quantum $XY$ model and the $H_f$ part being an interacting $f$-particle spinon part. We will then make a second mean-field treatment for each part of the Hamiltonian to convert them into effectively non-interacting ones. At half-filling, particle-hole symmetry requires the Lagrange multiplier $h =0$ and $\mu_0 = (U_0-U_2)/4$. We introduce three mean-fields as follows:

\begin{eqnarray}
\Delta = \frac{|V|}{2} \langle f^\dagger_{i,3/2} f^\dagger_{i,-3/2}-f^\dagger_{i,1/2} f^\dagger_{i,-1/2} \rangle_f
\end{eqnarray}

\begin{eqnarray}
Q_\theta = \sum_\sigma \langle f^\dagger_{i \sigma} f_{j \sigma} \rangle_f
\end{eqnarray}

\begin{eqnarray}
Q_f = \langle e^{i(\theta_i-\theta_j)} \rangle_\theta
\end{eqnarray}

\noindent where $i$ and $j$ are nearest-neighbor sites. The subscript $f$ or $\theta$ means that the quantum and thermal expectation values must be taken with respect to the spinon and roton sectors, respectively. Here we make the assumptions that these expectation values are real and independent of bond directions. One can relax these assumptions and treat the orbital current around a plaquette. After performing the decoupling scheme, the spinon and rotor part of the Hamiltonian becomes,

\begin{eqnarray}
H_f = -t Q_f \sum_{\langle ij \rangle, \sigma}(f^\dagger_{i\sigma}f_{j\sigma} + h. c) - \mu_0 \sum_{i \sigma} f^\dagger_{i \sigma} f_{i \sigma} \nonumber \\ + \Delta \sum_i(f^\dagger_{i,3/2} f^\dagger_{i,-3/2} - f^\dagger_{i,1/2} f^\dagger_{i,-1/2} + h. c)
\end{eqnarray}

\begin{eqnarray}
H_\theta = -t Q_\theta \sum_{\langle ij \rangle}(X^\dagger_{i}X_{j} + h. c) - \lambda \sum_{i} X^\dagger_{i} X_{i}\nonumber \\ -\frac{1}{2 U} \sum_i(i \partial _\tau X^\dagger_i)(-i\partial_\tau X_i),
\end{eqnarray}

\noindent where $X_i = e^{i \theta_i}$ and $\lambda$ is the Lagrange multiplier to impose the condition $|X_i|^2 = 1$. At the operator level, the Hamiltonian is now decoupled and while the spinon part is quadratic, the rotor part is naturally interacting. While the mean field parameter $Q_f$ renormalizes the hopping term and is related to the effective mass $m^\ast = m Q_f$, the expectation value of the pairing operator, $\Delta$ represents the pairing of spinons.

\section{V. Mean-Field treatment to spinon part}

The spinon part can easily be diagonalized in the momentum space. Performing a Fourier transform into momentum space and then the usual Bogoliubov transformation, the spinon Hamiltonian has the form,

\begin{eqnarray}
H_f = \sum_{k,l} \Lambda_l(k) \eta^\dagger_{k,l} \eta_{k,l} + \frac{1}{2} \sum_{k,l}[A^l_{kk} - \Lambda_l(k)],
\end{eqnarray}

\noindent where $\eta_{k,l}$ is a four component vector representing quasi-spinons and $\Lambda_l(k) = \pm \sqrt{\epsilon_k^2 + \Delta^2}$ (twice) are the eigenvalues with $l = 1,2,3,4$. Here $A^l_{k k^\prime}$ is a $4 \times 4$ diagonal matrix with the diagonal element $\epsilon_k = -Q_f \gamma_k - \mu_0$, where $\gamma_k = 2t \sum_\alpha \cos k_\alpha$ with $\alpha = x, y, z$ and the lattice momentum is re-scaled with the lattice constant. The quantum and thermal expectation values of Eq. (9) with respect to the Hamiltonian $H_f$ lead to the gap equation,

\begin{eqnarray}
\frac{4}{V} = -\frac{1}{N_s} \sum_k \frac{\tanh(\beta E_k/2)}{E_k},
\end{eqnarray}

\noindent where $E_k = \sqrt{\epsilon_k^2 + \Delta^2}$ with the total number of lattice sites $N_s$ and dimensionless inverse temperature $\beta$. Summing over nearest-neighbors and then calculating the expectation value in Eq. (10) with respect to $H_f$ gives,

\begin{eqnarray}
\eta t Q_\theta = \frac{1}{N_s} \sum_{k, \sigma} \gamma_k n_k
\end{eqnarray}

\noindent where $\eta$ is the number of nearest neighbors and the average occupation $n_k$ is given by,

\begin{eqnarray}
n_k = \frac{1}{2} - \frac{\epsilon_k}{2 E_k} \tanh(\beta E_k/2).
\end{eqnarray}

\noindent The two self-consistent equations derived in Eq. (15) and Eq. (16) must be solved with the Eq. (11) which can be written as $Q_f = \langle X^\dagger_iX_j \rangle_\theta$.

\section{VI. Mean-Field treatment to roton part}

The calculation of $Q_f$ requires a special attention as $X$ bosons can undergo Bose-Einstein condensation. The metallic (or band-insulating) phase corresponds to the ordering of rotors, and thus spontaneously break the $O(2)$ symmetry. The rotor disordered phase corresponds to the Mott-insulating phase given that the system is non-superfluid. Note the the metal to Mott-insulator transition is driven by spontaneous global $U(1)$ symmetry associated with the charge degrees of freedom. However, the Mott transition emerging from the slave rotor approach does not break any spin rotational symmetry, thus the transition is into a non-magnetic phase.

The final self-consistent equation can easily be calculated using a functional integral approach to the roton part of the Hamiltonian with the constraint equation $|X_i|^2 = 1$. Introducing the rotor Green's function $G_\theta(k, \tau) =  \langle X_k(\tau) X^\dagger_k(0) \rangle$, the constraint equation becomes,

\begin{eqnarray}
\frac{1}{N_s} \sum_k \frac{1}{\beta} \sum_n G_\theta (k, i\nu_n) = 1,
\end{eqnarray}

\noindent where $\nu_n = 2n\pi/\beta$ are the bosonic Matsubara frequencies. In coherent state path integral representation, the rotor Green's function can be written as,

\begin{eqnarray}
G_\theta(k, \tau) =  \frac{\int \prod_{ki}^{} \frac{dX_{ki} dX^\ast_{ki}}{2 \pi i} X(\tau) X^\ast_k(0) e^{-S_\theta}}{\int \prod_{ki}^{} \frac{dX_{ki} dX^\ast_{ki}}{2 \pi i}e^{-S_\theta}},
\end{eqnarray}

\noindent where the time index $i$ labeling runs from $0$ to $\infty$ corresponding to $\tau = 0$ and $\tau = \beta$, respectively. The action in the momentum space associated with the rotor part of the Hamiltonian is given by,

\begin{eqnarray}
S_\theta = \int_{0}^{\beta} d\tau \sum_k X^\ast_k(-\frac{1}{2U} \partial^2_\tau - \lambda - Q_\theta \gamma_k) X_k.
\end{eqnarray}

\noindent Following the standard path integral formalism, the rotor Green's function for the non-zero wave vector is given by,

\begin{eqnarray}
G_\theta (k, i\nu_n) = [\nu_n^2/U + \lambda - Q_\theta \gamma_k]^{-1}.
\end{eqnarray}

\noindent Note that following Ref.~\cite{florens}, a renormalization of $U \rightarrow U/2$ has been performed to preserve the exact atomic limit. Then writing,

\begin{eqnarray}
\frac{1}{\beta} \sum_n G_\theta (k, i\nu_n) = \frac{U}{\beta} \sum_n \frac{1}{i \nu_n + \sqrt{U(\lambda - Q_\theta \gamma_k)}} \nonumber \\ \times \frac{1}{-i \nu_n + \sqrt{U(\lambda - Q_\theta \gamma_k)}},
\end{eqnarray}

\noindent and performing a suitable contour integration, we find

\begin{eqnarray}
\frac{1}{\beta} \sum_n G_\theta (k, i\nu_n) = \frac{U}{2 \sqrt{U(\lambda - Q_\theta \gamma_k})} \nonumber \\ \coth [\frac{\beta}{2} \sqrt{U(\lambda - Q_\theta \gamma_k)}].
\end{eqnarray}

\noindent Combining this with Eq. (18) and separating the $k = 0$ term leads to the constraint equation,

\begin{eqnarray}
1 = Z + \frac{1}{2 N_s} \sum_{k \neq 0} \sqrt{\frac{U}{\lambda - Q_\theta \gamma_k}} \nonumber \\ \times \coth [\frac{\beta}{2} \sqrt{U(\lambda - Q_\theta \gamma_k)}],
\end{eqnarray}

\noindent where $ 0 \leq Z \leq 1$ is the rotor condensate amplitude which represents the quasiparticle weight. As the rotor condensation indicates the transition into the metallic phase, a non-zero quasiparticle weight $Z$ represents the metallic state. In the non-interacting limit $Z \rightarrow 1$. Finally, summing over nearest-neighbors of Eq. (11) and transforming into Fourier space leads to

\begin{eqnarray}
\eta tQ_f = \eta t Z +\frac{1}{N_s}\sum_{k\neq 0} \frac{\gamma_k}{\beta} \sum_n G_\theta (k, i\nu_n).
\end{eqnarray}

\noindent Completing the contour integration, our final self consistent equation becomes,

\begin{eqnarray}
\eta tQ_f = \eta t Z -\frac{1}{2N_s}\sum_{k\neq 0} \gamma_k \sqrt{\frac{U}{\lambda - Q_\theta \gamma_k}} \nonumber \\ \times \coth [\frac{\beta}{2} \sqrt{U(\lambda - Q_\theta \gamma_k)}].
\end{eqnarray}

\noindent This $Q_f$ is the mass enhancement factor of the quasiparticle, thus it is proportional to the effective mass of the quasiparticle $m^\ast = Q_f m$, where $m$ is the bare mass of the free atoms. As the second term in Eq. (26) is negative, mass enhancement is always greater than the quasiparticle weight, $Q_f > Z$ at the saddle point level, and remains finite even away from the metallic phase where $Z$ vanishes.

\section{VII. Zero temperature formalism and quantum phase transitions}

For spin-3/2 atoms on a d-dimensional lattice at zero temperature, four self-consistent equations can be largely simplified. First, by introducing the d-dimensional density of states, $D(\epsilon) = \frac{1}{N_s} \int \frac{d^d k}{(2 \pi)^d} \delta(\epsilon + \gamma_k)$ and setting energy units to be $t =1$, our self consistent equations become,

\begin{eqnarray}
\frac{4}{V} = -\int d\epsilon D(\epsilon) \frac{1}{\sqrt{(Q_f \epsilon-\mu_0)^2 + \Delta^2}},
\end{eqnarray}

\begin{eqnarray}
\eta  Q_\theta = -\int d\epsilon D(\epsilon)\epsilon \biggr(\frac{1}{2} - \frac{Q_f \epsilon-\mu_0}{2 \sqrt{(Q_f \epsilon-\mu_0)^2 + \Delta^2}}\biggr),
\end{eqnarray}

\begin{eqnarray}
1 = Z + \frac{1}{2} \int d\epsilon D(\epsilon) \sqrt{\frac{U}{\lambda + Q_\theta \epsilon}},
\end{eqnarray}

\noindent and,

\begin{eqnarray}
\eta  Q_f = \eta  Z - \frac{1}{2} \int d\epsilon D(\epsilon) \epsilon \sqrt{\frac{U}{\lambda + Q_\theta \epsilon}},
\end{eqnarray}

\noindent where the nearest-neighbor coordination number is $\eta$ and these self-consistent equations are valid only at zero temperature for a half filled system whose chemical potential is given by $\mu_0 = (U_0 - U_2)/4$.

Obviously, the superconducting phase is characterized by the non-zero singlet pairing order parameter $\Delta$. The gap equation gives non-zero solutions for the pairing order parameter for all $V < 0$, thus the superconducting transition line in $U_0-U_2$ parameter space is given by the equation $U_0 = U_2$. This is the SU(4) symmetric line which can be alternatively represented by $\mu_0 = 0$. In the metallic phase rotors are condensed so that the non-zero value of the condensate amplitude or the quasiparticle weight $Z$ signifies the metallic state. In the metallic phase, a macroscopic fraction of rotors occupies the lowest energy $E_{l} = - \eta t Q_\theta$ and the Lagrange multiplier or the rotor chemical potential $\lambda = -E_l \equiv \eta tQ_\theta$ remains constant. The quantum phase transition from the metallic state to the Mott-insulating state is characterized by the vanishing quasiparticle weight $Z$. In the Mott-insulating phase the quasiparticle weight $Z$ is zero and the rotor chemical potential $\lambda > \eta tQ_\theta$ needs to be determined self consistently. The metal-insulator transition line can be determined by setting $\Delta =0$, $Z = 0$, and $\lambda = \eta t Q_\theta$ in self-consistent equations presented above.

For a two-dimensional square lattice, the density of states can be approximated by a closed form using the elliptic integral of the first kind $K$, $D(\epsilon) = \frac{1}{2 \pi^2}K(1-\epsilon^2/16)$ for $ -4t \leq \epsilon \leq 4t$, and zero otherwise. Therefore for both metal and Mott-insulating phases, where $\Delta = 0$, $Q_\theta$ has an analytical form. Evaluating the integral in Eq. (28) for a two-dimensional square lattice, we find,

\begin{eqnarray}
Q_\theta = \frac{1}{2 \pi^2} K(1 - \mu_0^2/16) [16 - \mu_0^2].
\end{eqnarray}

\noindent Notice that the chemical potential $\mu_0 = (U_0 - U_2)/4$ at half filling, thus $Q_\theta$ depends on the interaction parameters. This is in contrast with the regular Hubbard model with spin-1/2 particles where $Q_\theta = 4/\pi^2$ is independent of the interaction~\cite{SG}. Here at the SU(4) limit where $\mu_0 = 0$, we find $Q_\theta = 8/\pi^2$ where the extra factor 2 comes from the extra spin for the spin-3/2 system. For a three-dimensional cubic lattice, we numerically evaluate the $Q_\theta$ value. Re-arranging Eq. (30), we find a self-consistent equation for the critical $U_2$ value for the metal-insulator transition,

\begin{eqnarray}
U_2^{CMI} = 16 Q_\theta^C \biggr\{ \int \frac{d\epsilon D(\epsilon)}{\sqrt{1 + \epsilon/4}} \biggr\}^{-2},
\end{eqnarray}

\noindent where $Q_\theta^C = Q_\theta (U_2 = U_2^{CMI})$ depends on the critical interaction through the chemical potential.

\begin{figure*}
\includegraphics[width=\columnwidth]{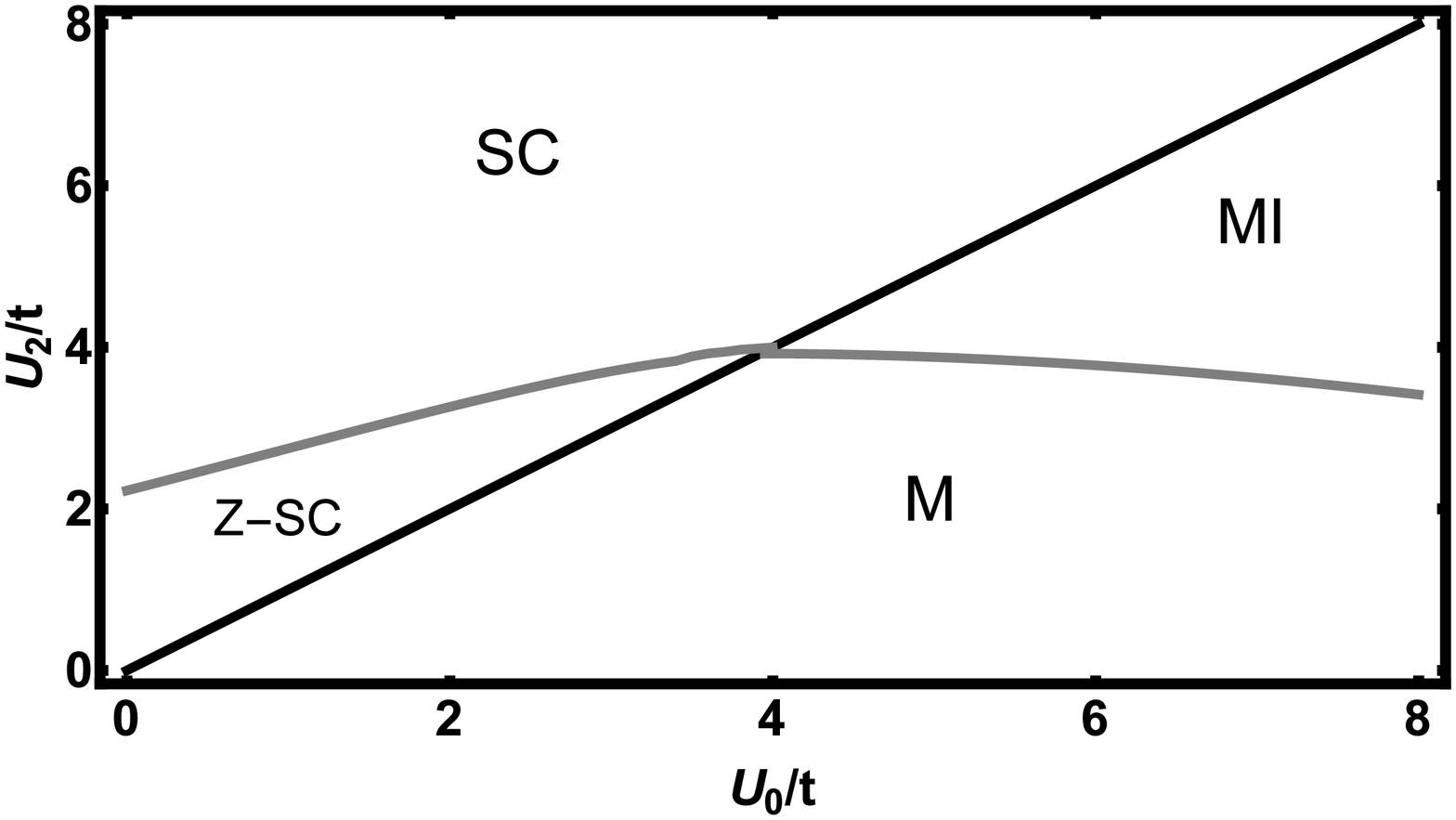}\hfill
\includegraphics[width=\columnwidth]{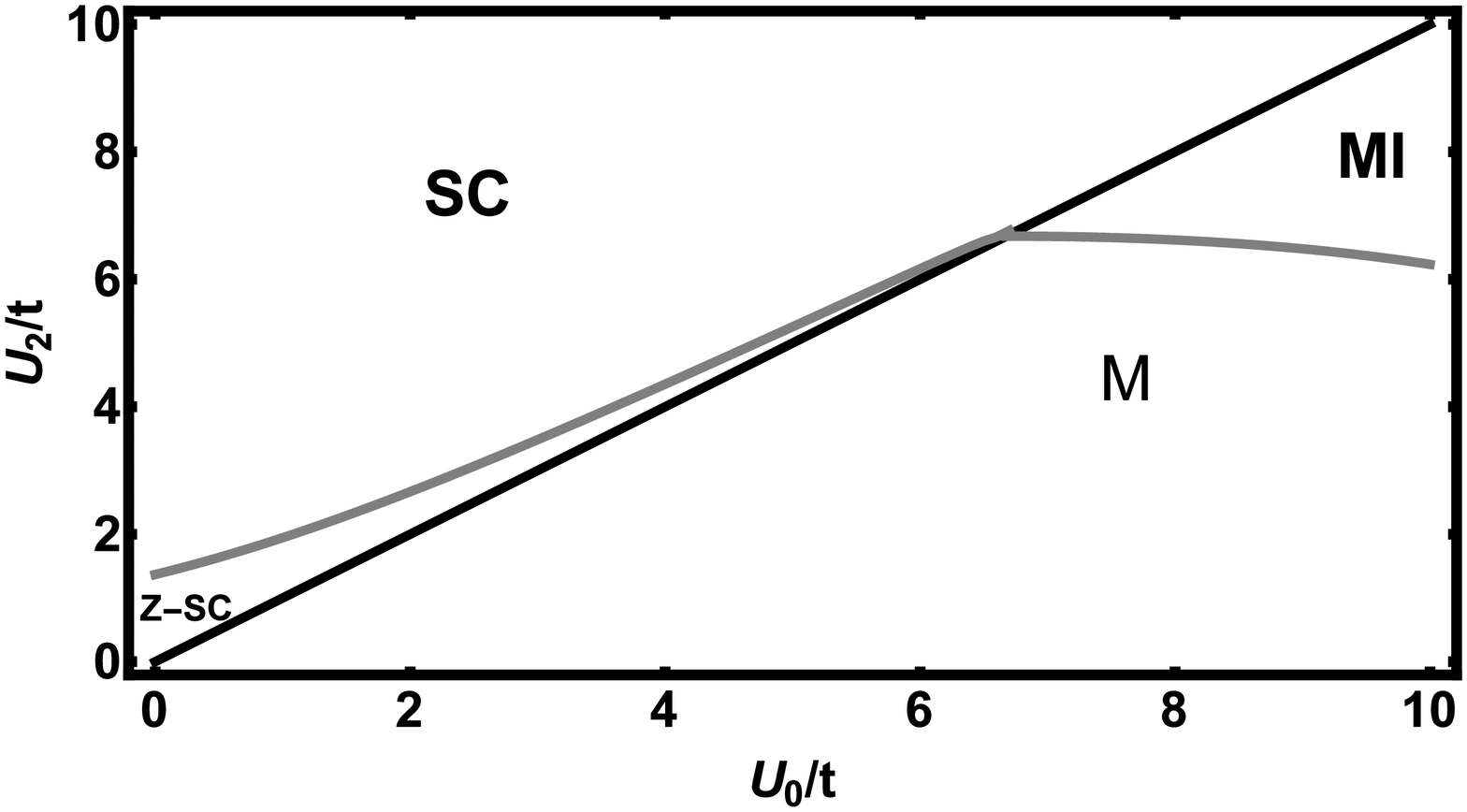}
\caption{Zero temperature phase diagram showing four different phases; M: metallic phase, MI: Mott-insulating phase, Z-SC: superconducting phase with non-zero quasiparticle weight, and SC: superconducting phase with zero quasiparticle weight. Left: Phase diagram for the two-dimensional square lattice. Right: Phase diagram for the three-dimensional cubic lattice. The metal phase is characterized by a global U(1) symmetry broken state associated with the rotor degrees of freedom and both superconducting phases (SC and Z-SC) are characterized by the global U(1) symmetry broken states associated with the spinon degrees of freedom. The Z-SC phase shows additional U(1) symmetry breaking associates with the rotor sector.}\label{ZTPD}
\end{figure*}

Zero temperature phase diagrams for both square and cubic lattices in the $U_0 - U_2$ plane are given in FIG.~\ref{ZTPD}. There are four different phases at half filling, a metallic phase (M) at smaller value of $U_2 < U_0$, a Mott-insulating phase (MI) at larger values of $U_2 < U_0$, and two distinct superconducting phases (Z-SC and SC) for larger values of $U_2 > U_0$. In the Z-SC phase quasiparticle weight $Z$ is non-zero and it reaches zero at the SC - Z-SC boundary breaking the O(2) rotor symmetry. The boundary of the Z-SC and SC phases is determined by solving our self-consistent equations with the conditions, $Z =0$ and $\lambda = \eta t Q_\theta$. With our numerical calculations, we find that $Q_f$ remains constant along the Z-SC-SC boundary, giving $Q_f \approx 0.255$ for a two-dimensional square lattice and $Q_f \approx 0.104$ for a three-dimensional cubic lattice. As to our knowledge, this new emergent Z-SC phase in which both U(1) symmetries associated with the rotor degrees of freedom and spinon degrees of freedom are broken, has not been discussed before.

 \begin{figure*}
\includegraphics[width=\columnwidth]{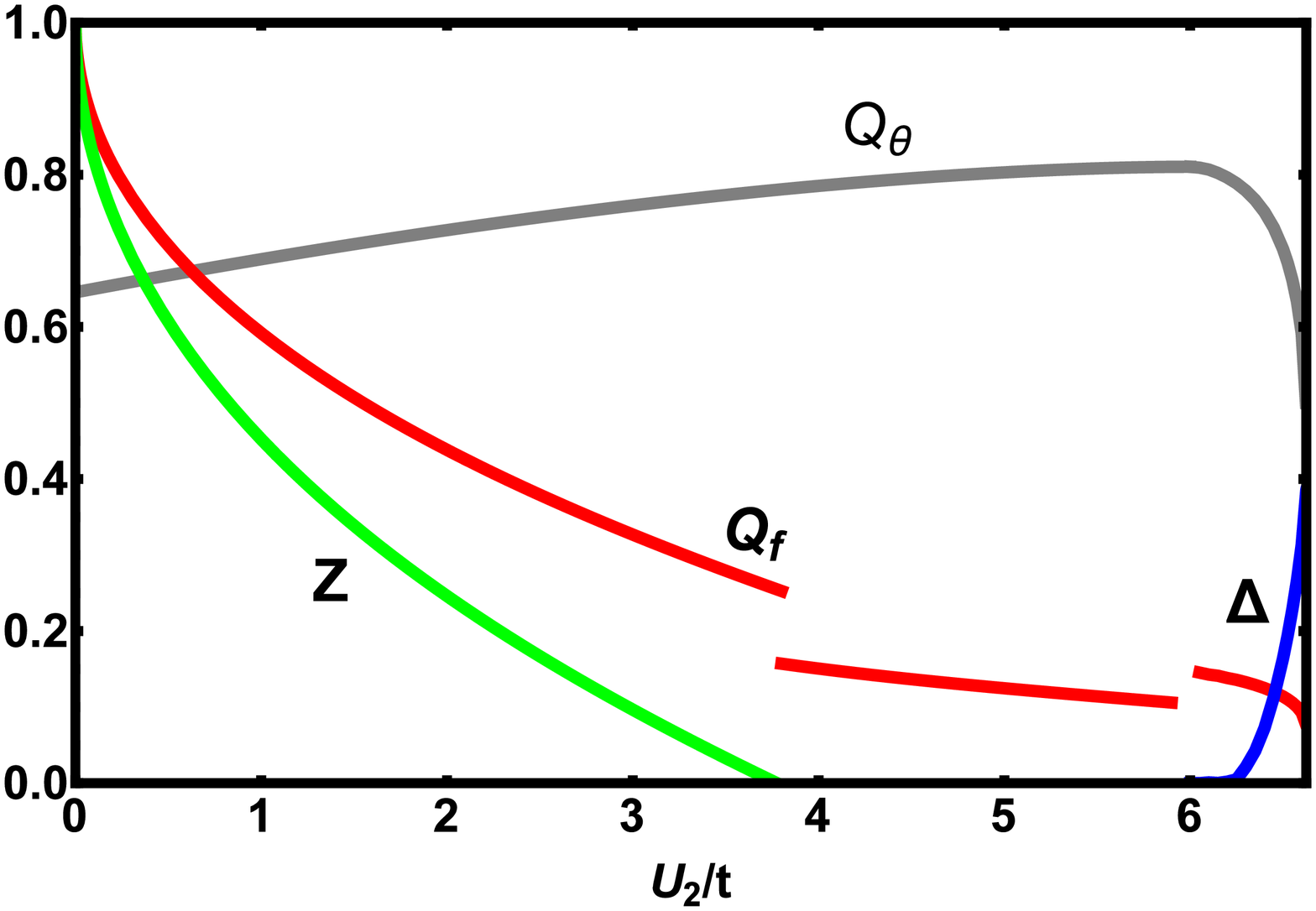} \hfill
\includegraphics[width=\columnwidth]{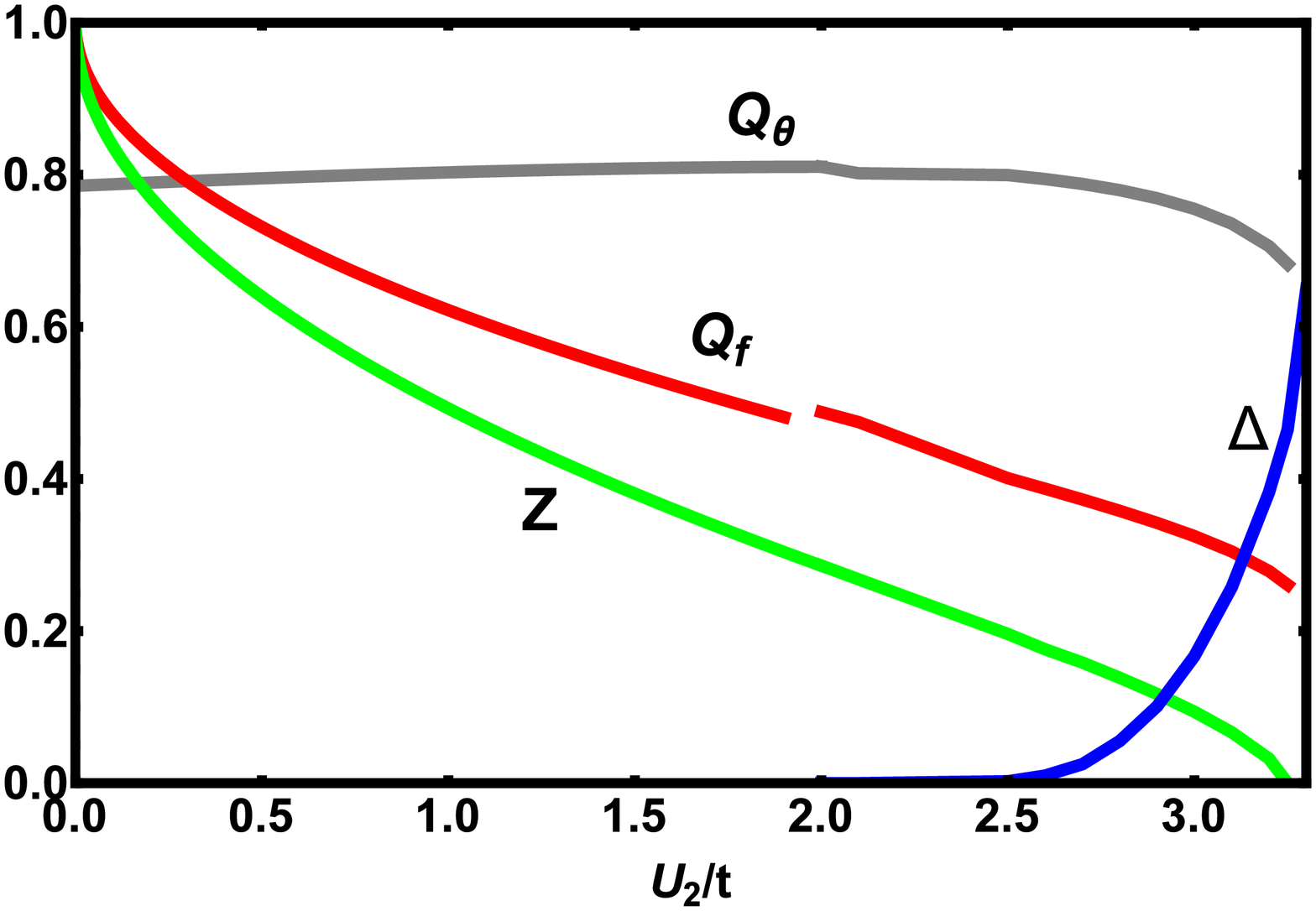}
\caption{(color online) Zero temperature physical properties as a function of $U_2$ at $U_0 = 6 t$ (Left Panel) and at $U_0 = 2 t$ (Right Panel). \emph{Left}: The quasiparticle weight $Z$ (green line) continuously and monotonically decreases from unity to zero showing metal to Mott-insulator transition at $U_2 \approx 3.8 t$. For $3.8 t \leq U_2 \leq 6 t$, both the quasiparticle weight and the pairing order parameter remain zero showing the Mott-insulating phase. For $U_2 > 6 t$, the singlet pairing order parameter $\Delta$ (blue line) becomes finite, indicating the Mott-insulator to superconductor transition. \emph{Right}: Only the metallic phase exists for low values of $U_2$ showing non-zero $Z$ and zero $\Delta$. For $2 t \leq U_2 \leq 3.3 t$, the existence of the Z-SC phase is evident as both $Z$ and $\Delta$ have non-zero values. Notice for both cases that, while $Q_\theta$ (gray line) is continuous at each phase boundary, $Q_f$ shows discontinuities. For clarity, the pairing order parameter $\Delta$ has been increased by a factor five in these figures.}\label{OPAT}
\end{figure*}

For the two-dimensional square lattice, the quasiparticle weight $Z$, the mass enhancement $Q_f$, the average kinetic energy of the spinons $Q_\theta$, and the pairing order parameter $\Delta$ are shown at fixed values of $U_0 =6 t$ and $U_0 = 2 t$ in Fig.~\ref{OPAT}. As can be seen from the left panel of FIG~\ref{OPAT}, the quasiparticle weight is unity at a non-interacting level of rotors, and then reaches zero at the metal-insulator boundary. Meantime the superconducting order parameter picks a finite value at the insulator-superconductor boundary and increases as one increases the interaction parameter $U_2$ beyond $U_0$. Both the quasiparticle weight and superconducting order parameter remain at zero in the intermediate Mott-insulating phase. The right panel of FIG~\ref{OPAT} shows the variation of parameters across the quantum phase transition from metallic to Z-SC to SC phases. For a fixed value of $U_0 = 2 t$, the metallic phase exists for low values of $U_2 \leq 2$ indicating a non-zero quasiparticle weight $Z$ and zero pairing order parameter $\Delta$. For intermediate values of $U_2$, both the quasi particle weight and pairing order parameter become non-zero, and thus represents the Z-SC phase. As can be seen from the right panel in the SC phase, while the quasiparticle weight vanishes, the pairing order parameter remains non-zero beyond $U_2 \approx 3.3 t$. Notice, that both the mass enhancement and average kinetic energy of the rotons are non-zero across all quantum phase transitions, however $Q_f$ shows small discontinuities at the quantum phase transitions. This zero-temperature discontinuity of the mass enhancement factor $Q_f$ is an artifact of the mean-field theory and it can be recovered by adding fluctuations over the mean fields, as we discussed in discussion section.

\section{VIII. Finite temperature phase transitions}

\begin{figure*}
\includegraphics[width=\columnwidth]{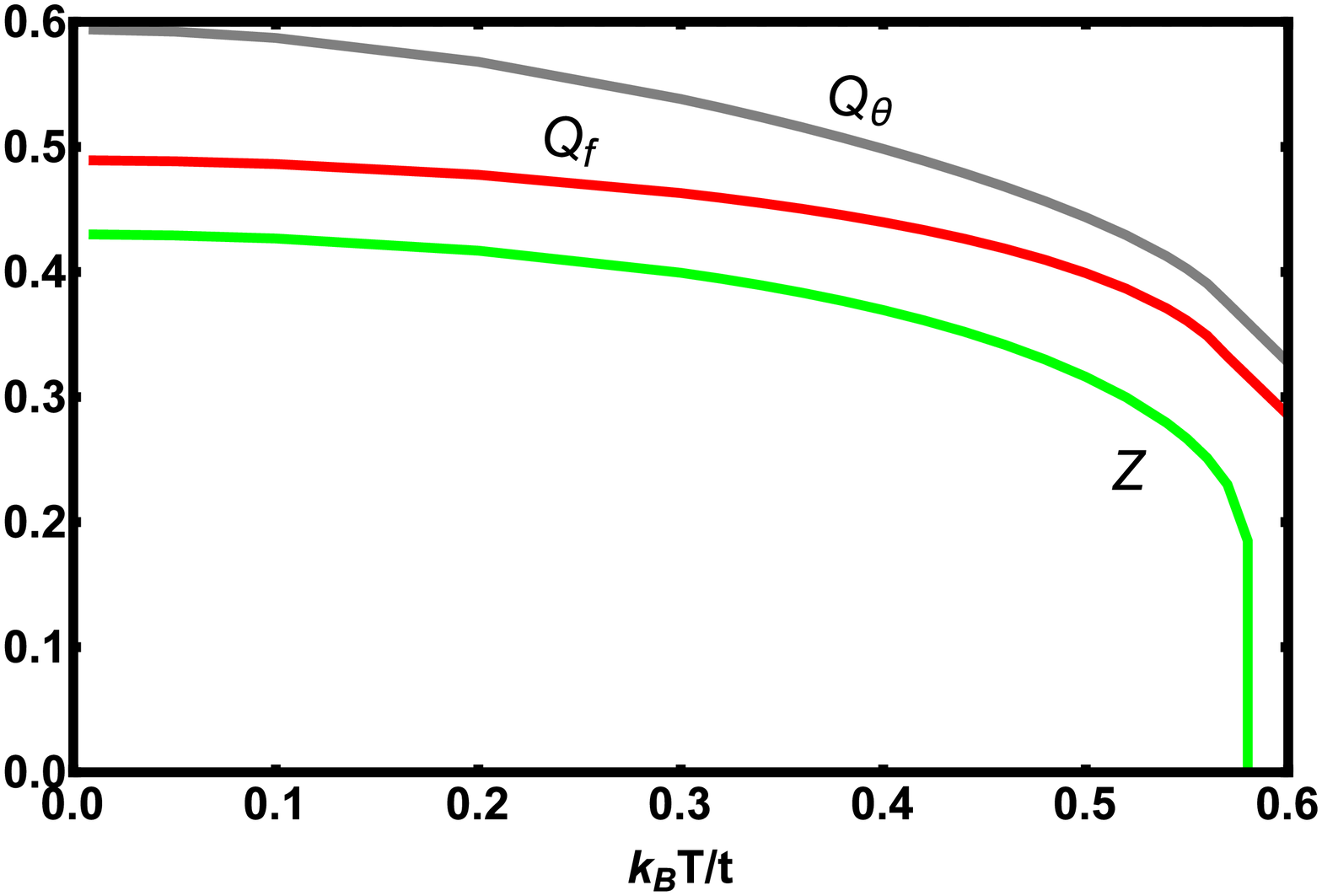}\hfill
\includegraphics[width=\columnwidth]{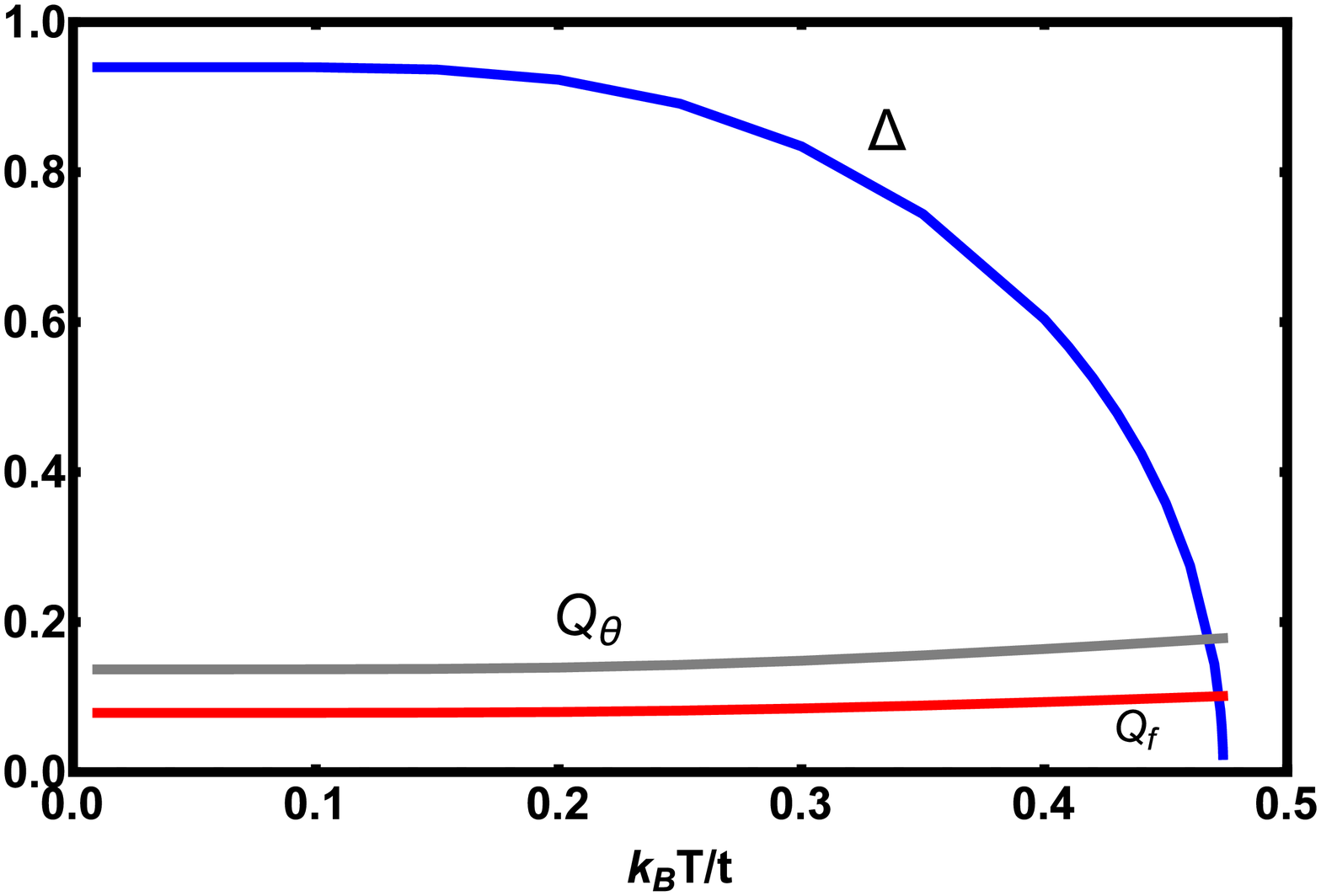}
\caption{(color online) Finite temperature properties of the Z-SC phase (right) and metallic phase (left) of fermions in three dimensional cubic lattice. The left graph shows the quasiparticle weight ($Z$) as a function of temperature $k_BT/t$ in the metallic phase. The interaction parameters were chosen as $U_2 = 2 t$ and $U_0 = 4 t$. The quasiparticle weight (green line) monotonically decreases from the ground state metallic phase to the Mott-insulating phase at $k_BT/t \approx 0.58 t$. Notice that the finite temperature metal-insulator transition is first order showing a discontinuity of $Z$ at the transition. At the transition, both $Q_\theta$ and $Q_f$ are finite and continuous. The right graph shows the superconducting order parameter ($\Delta$) as a function of temperature $k_BT/t$ in the Z-SC phase close to the Z-SC-SC boundary. The interaction parameters were chosen as $U_2 = 1.5 t$ and $U_0 = 0.5 t$,  While $\Delta$ continuously decreases (blue line) from a finite value to zero showing a second order thermal phase transition, both $Q_f$ and $Q_\theta$ values remain almost constant as a function of temperature.}\label{FTMIT}
\end{figure*}

For two-dimensional fermions on a lattice, the finite temperature phase transitions are absent, but one can expect to have crossovers. For three-dimensional fermions on a lattice, the finite temperature phase transitions are not forbidden. We numerically solve the finite temperature self consistent equations for the cubic lattice. As a demonstration, we show some finite temperature properties of the metallic phase and Z-SC phase in FIG.~\ref{FTMIT}. The left panel shows the temperature dependence of the quasiparticle weight $Z$, the mass enhancement factor $Q_f$, and, the average kinetic energy $Q_\theta$ in the metallic phase where the interaction parameters are fixed at $U_2 = 2t$ and $U_0 = 4t$. Unlike the zero temperature metal-Mott insulator transition, the finite temperature metal-insulator phase transition is found to be of first order, thus $Z$ shows a discontinuity at the transition. However, we find that $Q_f$ and $Q_\theta$ remain to be continuous at the transition. The right panel shows the temperature dependence of the singlet pairing order parameters $\Delta$, $Q_f$ and $Q_\theta$ for the interaction parameters $U_2 = 1.5 t$ and $U_0 = 0.5 t$. The ground state for these interaction parameters is the Z-SC phase where both the singlet pairing order parameter $\Delta$ and the quasiparticle weight $Z$ are non-zero. The continuously vanishing singlet order parameter at a high temperature indicates the second order thermal transition from the Z-SC phase.

\section{IX. Discussion and conclusions}

As discussed above, all zero temperature quantum phase transitions are second order, however we find a discontinuity in the mass enhancement factor $Q_f$ when the transition is into a Mott-insulating phase. On the other hand, while the finite temperature thermal phase transition into the Mott-insulating phase is first order, we do not find any discontinuity in the mass enhancement factor. In our approach, the zero temperature Mott-transition is continuous, thus one expects continuous destruction of the metallic Fermi surface. This would lead to a metal having instabilities toward a magnetic Mott-insulator due to the Fermi surface nesting, though we have not considered any symmetry breaking insulating states in our approach.

The two superconducting phases discussed above are distinguished for two main reasons. In the Z-SC phase, both the rotor and the pair of spinons are in the condensate. As a result, the broken U(1) gauge symmetry in the rotor sector gives a non-zero quasiparticle weight similar to that of the metallic phase. In the SC phase, the rotors give a non-zero charge gap $\delta_c = 2\sqrt{U(\lambda-\eta t Q_\theta)}$, similar to that of the Mott-insulating phase.

In the present work, we have decoupled the rotor and spinon parts of the Hamiltonian using a mean-field theory. We do not expect the inclusion of fluctuation to alter the qualitative features. However the physical observable can be slightly different once the direct coupling between the rotors and spinons is restored. Fluctuations can easily be included by going beyond the saddle point approximation. From the transition from the metal phase or the Z-SC phase, the quasiparticle weight vanishes, however the effective spinon hopping $tQ_f$ is finite. As a result, the effective mass does not diverge at these transitions. We believe this is an artifact of our mean-field theory. In the presence of fluctuation of a gauge field, the zero-sound Goldstone mode will combine with a gauge boson through the Anderson-Higgs mechanism. We believe that this would recover our metal phase as a proper Fermi liquid phase with a diverging Fermi liquid mass~\cite{WWK}. In addition to the mean-field approximation, we treat our constraint globally and assume that all parameters are bond independent. We do not expect these approximations to change any qualitative features, specially for the square and cubic lattices discussed.

On the experimental side, spin-3/2 alkaline-earth atoms, such as $^{135}$Ba and $^{137}$Ba can be promising candidates for observing the Z-SC novel emerging phase. Even though the full spectrum of scattering lengths is not available yet, it is predicted that both scattering lengths $a_0$ and $a_2$ should have similar values~\cite{R2, zsc1}.  Therefore, we believe experiments can find a suitable parameter window in the phase diagram to observe the Z-SC phase even if scattering lengths cannot be independently tuned. On the other hand, the pairing phenomena and the emerging novel phase discussed here are much more general concepts associated with many body systems. Thus, the Z-SC phase must exist in other many body systems where competing interactions are taking place. One such example is spin-3/2 rare-earth based half-Heusler semimetals~\cite{ zsc2, zsc3, zsc4, zsc5,zsc6, zsc7}. Another promising electronic compound is rubidium doped fullerids~\cite{ zsc8}. Though on-site interaction is repulsive, the effective negative Hund coupling due to phonon screening effects can lead to a pairing of electrons~\cite{ zsc9, zsc10}.  Indeed, a recent experiment finds a novel phase refered to as a Jahn-Teller metal in rubidium doped fullerids~\cite{zsc8}. Perhaps, the superconducting critical temperature enhancement close to the tri-critical point of a paramagnetic metal, a paramagnetic insulator, and a superconducting phases indicates the Z-SC phase in  this rubidium doped fullerid compound~\cite{zsc11}.

The superfluid density, charge gap, and quasiparticle weight all can be measured with currently available experimental techniques in cold gas experiments. For example, the momentum distribution of the atoms can be probed by the absorption imaging after a period of ballistic expansion or in trap in-situ imaging~\cite{FDop4}. The charge gap can be detected by measuring the fraction of atoms residing in a lattice site~\cite{FDop3}. The superconducting order can be probed by the momentum-resolved Bragg spectroscopy~\cite{SForder}. In addition, the periodic forcing can also be used as a detection and manipulation tool for many-body states of ultra-cold atomic quantum gases in optical lattices~\cite{PEF}. Though we neglected it in this study, the underlying harmonic trapping potential present in all cold gas experiments causes the density to monotonically vary across the lattice. As a result, the edge of the trap will not be at half filling. As the metallic phase is favorable over the Mott-insulating phase away from half filling, we expect the metallic phase to dominate over the Mott-insulating phase in the phase diagram.

In conclusion, we have studied the competition of emerging phases of spin-3/2 fermions subjected to a periodic lattice potential using a slave rotor approach. In addition to the well known Fermi liquid metallic phase, Mott-insulating phase, and singlet pairing superconducting phase, we discovered the possibility of having a novel emerging superconducting phase due the competing interactions. The novel superconducting phase is characterized by the global U(1) broken symmetries with respect to both roton and spinon fields. Experimentally, this novel phase can be differentiated from the regular superconducting phase by its non-zero quasiparticle weight. Further, we have calculated properties of these phases for fermions in both a two dimensional square lattice and a three dimensional cubic lattice geometries at zero and finite temperatures.

\section{X. Acknowledgments}
The author acknowledges the support of Augusta University, the hospitality of ITAMP at the Harvard-Smithsonian
Center for Astrophysics and KITP at UC-Santa Barbara. The initial part of the research was undertaken at ITAMP and ITAMP is supported by a grant from the National Science Foundation to Harvard University and the Smithsonian Astrophysical Observatory. The research was completed at KITP and was supported in part by the National Science Foundation under Grant No. NSF PHY11-25915.

\end{document}